\documentclass[journal]{IEEEtran}[12pt]
\usepackage[table]{xcolor}
\usepackage{psfrag}
\usepackage{acronym}
\usepackage{etoolbox}
\usepackage{wrapfig}
\usepackage{epsfig}
\usepackage{colortbl,color}
\usepackage{amsmath}
\usepackage{amssymb}
\usepackage{mathabx}
\usepackage{enumerate}
\usepackage{bbm}
\usepackage{algorithm}
\usepackage{algorithmic}
\usepackage{lipsum,amsmath,multicol}
\usepackage{stfloats}
\usepackage{arydshln} 
\usepackage{amsfonts}
\usepackage{dsfont}
\usepackage{upgreek}
\usepackage{pgfplots} 
\usepackage{soul}
\usepackage{tabularx}
\usepackage{multirow}
\usepackage{graphicx}
\usepackage{pgfplots}
\usepackage{url}

\usepackage{cite}
\usepackage{verbatim}
\acrodef{1G}{first generation}
\acrodef{2G}{second generation}
\acrodef{3G}{third generation}
\acrodef{4G}{fourth generation}
\acrodef{5G}{fifth generation}
\acrodef{6G}{sixth generation}
\acrodef{AI}{artificial intelligence}
\acrodef{CPL}{central perpendicular line}
\acrodef{BS}[BS]{base station}
\acrodef{CRLB}{Cramer-Rao lower bound}
\acrodef{CS}{compressive sensing}
\acrodef{CID}{cellular ID}
\acrodef{DCS}{digitally controllable scatterer}
\acrodef{DMA}{dynamic metasurface antenna}
\acrodef{EM}{electromagnetic}
\acrodef{FIM}{Fisher information matrix}
\acrodef{KKT}{Karush–Kuhn–Tucker}
\acrodef{KPI}{key performance indicator}
\acrodef{LOS}{line-of-sight}
\acrodef{CSI}{channel state information}
\acrodef{EMF}{electromagnetic fields}
\acrodef{E-CID}{enhanced-CID}
\acrodef{ES}{emergency services}
\acrodef{IoT}{internet-of-things}
\acrodef{LIS}{large intelligent surface}
\acrodef{LPP}{LTE Positioning Protocol}
\acrodef{mm-waves}[MM-Waves]{millimeter-waves}
\acrodef{MLE}{maximum likelihood estimator}
\acrodef{SISO}{single-input single-output} 
\acrodef{MAP}{maximum a posteriori} 
\acrodef{MIMO}{multiple-input multiple-output}
\acrodef{MS}[UE]{user equipment}
\acrodef{Multi-RTT}{multi round trip time}
\acrodef{NLOS}{non-line-of-sight}
\acrodef{NR}{new radio}
\acrodef{NRPPa}{NR Positioning Protocol A}
\acrodef{OFDM}{orthogonal frequency-division multiplexing}
\acrodef{RFID}{radio-frequency identification}
\acrodef{RIS}{reconfigurable intelligent surface}
\acrodef{RSS}{received signal strength}
\acrodef{SDS}{software defined surface}
\acrodef{SRE}{smart radio environment}
\acrodef{SNR}{signal-to-noise ratio}
\acrodef{SVD}{singular value decomposition}
\acrodef{SL}{service level}
\acrodef{POA}{phase-of-arrival}
\acrodef{TOA}{time-of-arrival}
\acrodef{TDOA}{time difference-of-arrival}
\acrodef{COA}{curvature-of-arrival}
\acrodef{AOA}{angle-of-arrival}
\acrodef{AOD}{angle-of-departure}
\acrodef{RSSI}{received signal strength indicator}
\acrodef{SLAM}{simultaneous localization and mapping}
\acrodef{CRLB}{Cramér-Rao lower bound}
\acrodef{GDOP}{geometric dilution of precision}
\acrodef{PEB}{position error bound}
\acrodef{pdf}{probability density function}
\acrodef{PE}{positioning element}
\acrodef{PDOA}{phase difference-of-arrival}
\acrodef{IPS}{smartphone-centered indoor positioning systems}
\acrodef{PRS}{position reference signals}
\acrodef{OEB}{orientation error bound}
\acrodef{RMSE}{root mean square error}
\acrodef{UAV}{unmanned aerial vehicle}
\acrodef{UTDOA}{uplink time difference-of-arrival}
\acrodef{UAoA}{uplink angle-of-arrival}
\acrodef{eMBB}{enhanced mobile broadband}
\acrodef{RA}{resource allocation}
\acrodef{RF}{radio frequency}
\acrodef{QoS}{quality of service}
\acrodef{ESPRIT}{Estimation of Signal Parameters via Rotational Invariance Technique}
\acrodef{MUSIC}{Multiple Signal Classification}
\acrodef{LS}{least squares}
\acrodef{LCS}{location service}
\acrodef{E-OTDOA}{enhanced observed time difference-of-arrival}
\acrodef{ELAA}{extremely large aperture arrays}
\acrodef{OTDOA}{observed time difference-of-arrival}
\acrodef{GPS}{global positioning system}
\acrodef{GSM}{global system for mobile communications}
\acrodef{PRS}{position reference signals}
\acrodef{UE}{user equipment}
\acrodef{UWB}{ultra-wide bandwidth}
\acrodef{XR}{Extended Reality}
\acrodef{VLC}{visible light communication}
\acrodef{VLP}{visible light positioning}
\acrodef{LiDAR}[Li-DAR]{light detection and ranging}
\acrodef{UHF}{ultra high frequency}

\title{Towards 6G Holographic Localization: \\ Enabling Technologies and Perspectives}
\author{Ahmed~Elzanaty,
~\IEEEmembership{Senior Member,~IEEE}, 
Anna Guerra,
~\IEEEmembership{Member,~IEEE,} ,
Francesco Guidi,
~\IEEEmembership{Member,~IEEE,} \\
Davide Dardari,
~\IEEEmembership{Senior Member,~IEEE,}
 and Mohamed-Slim Alouini,
~\IEEEmembership{Fellow,~IEEE}
\thanks{A. Elzanaty is with the Institute for Communication Systems (ICS), Home of 5G\&6G Innovation Centres (5G\&6GIC), University of Surrey, Guildford, GU2 7XH, United Kingdom, email: a.elzanaty@surrey.ac.uk.}
\thanks{A.G. and D.D. are with the DEI/WiLAB-CNIT, University of Bologna, Italy, email: \{anna.guerra3,davide.dardari\}@unibo.it.}
\thanks{F.G. is with CNR-IEIIT, Italy, e-mail: francesco.guidi@ieiit.cnr.it.}
\thanks{M.-S. Alouini is with  King Abdullah University of Science and Technology, Thuwal 23955-6900, Saudi Arabia, e-mail: slim.alouini@kaust.edu.sa.}

}
\definecolor{aliceblue}{rgb}{0.94, 0.97, 1.0}
\definecolor{greenyellow}{rgb}{0.7, 0.9, 0.4}
\pgfplotsset{compat=1.16}

\begin{document}

\maketitle

\begin{abstract}
In the last years, we have experienced the evolution of wireless localization from being a simple add-on feature for enabling specific applications to become an essential characteristic of wireless cellular networks, as for \ac{6G} cellular networks. This paper illustrates the importance of radio localization and its role in all the cellular generations, from \ac{1G} to \ac{6G}. Also, it speculates about the idea of holographic localization where  the characteristics of  \ac{EM} waves, including the spherical wavefront in near-field, are fully controlled and exploited to achieve better wireless localization. Along this line, we briefly overview possible technologies, such as large intelligent surfaces, and challenges to realize holographic localization. 
To corroborate our vision, we also include a numerical example that confirms the potentialities of holographic localization. 
\end{abstract}

\begin{IEEEkeywords}
Holographic Localization; Large Intelligent Surface; Metasurfaces.
\end{IEEEkeywords}

\acresetall
\bstctlcite{IEEEexample:BSTcontrol}

\section{Introduction}
Beyond \ac{5G} cellular networks are endorsed by wide available bandwidths, high frequencies, and antenna arrays with a large number of antenna elements, i.e., massive \ac{MIMO}. These features will allow not only high-speed communication but also high-accuracy wireless positioning at an unprecedented scale \cite{HaddiAccess:21}.
For instance, massive \ac{MIMO} with considerably large distance between antenna elements (compared to the wavelength) can enable  \ac{ELAA}. These arrays have significantly high spatial resolution, allowing to attain accurate positioning performance due to their large apertures. \ac{ELAA} can be enabled by several technologies such as coordinated multipoint and cell-free massive \ac{MIMO}\cite{BJORNSONHenk:19}.

On the contrary,  we may be interested in increasing the antenna densification into compact sizes so that they can be integrated into small areas. Operating like this, in the case of extremely dense antennas, there will be the possibility to create a spatially (quasi-) continuous \ac{EM} aperture, i.e., a holographic \ac{MIMO} array \cite{Huang2020holographic}. A promising solution for the realization of a quasi-continuum antenna array is provided by intelligent surfaces, usually made of metamaterials \cite{hu2018beyond}.

In this paper, we envision holographic localization as the future of wireless positioning that will be characterized by considering  \ac{ELAA} and intelligent surfaces to exploit near-field and achieve the control of {EM} waves at unprecedented level.
The term {\em holographic} here intended as the ability to fully control and/or exploit the characteristics offered by different \ac{EM} propagation regimes obtained by an optimized control of the radiating beams and operating directly on the electric field.\footnote{Holographic localization is different from the traditional concept of {\em holography}, which instead concerns the practice of making holograms, e.g., photographic recordings of a light field to reproduce 3D images.} This is particularly important when using electrically large antennas at high frequency, e.g., the millimeter wave or THz. Hence, the operating conditions may easily fall in the Fresnel radiating near-field region, where the classical plane wave propagation assumption is no longer valid.
In this regime, the \ac{EM} wavefront curvature offers the possibility to associate more information about the location and the antenna orientation to the wave, allowing, if fully exploited,  improved localization performance. As an example, different propagation regimes of the \ac{EMF} can be exploited to infer the position information by controlling the \ac{EM} waves generated or sensed by antennas of multiple devices. Also, the features of a transmitting device, e.g., the device position and  array shape and orientation, have a significant impact on the received signal. Hence, these characteristics can be inferred from the signal by considering a model that accounts for the  equations governing the propagation of \ac{EM} waves, rather than  standard models \cite{Fri:19}. 

Hence, our vision is to \emph{fully control} and \emph{fully exploit} the wavefront to enhance the positioning performance by considering the spherical wavefront  and operating on the \ac{EM} level. Towards our vision, we first overview the history of wireless positioning, emphasizing the importance of holographic localization for new applications. Then, we discuss about possible enabling technologies to control the wavefront, including \acp{LIS}, and possible positioning algorithms together with ultimate localization to fully exploit the optimized wavefront. Finally, as an use case, we assess the performance of a \ac{RIS}-aided localization system that optimizes the \ac{RIS} phases to  minimize the positioning error, and make use of the spherical wavefront through \ac{MLE} considering the near-field model. 

\section{Localization History and New Directions}
Wireless localization has been an essential service for several applications in the last decades \cite{win2018theoretical}. In general, the position of a \ac{UE} can be estimated either at the \ac{UE} ({\em downlink positioning})  or at the network-level ({\em uplink positioning}).  
Different types of measurements can be considered for positioning, e.g., \ac{TOA} of the radio signal, \ac{UTDOA}, \ac{TDOA}, \ac{PDOA}, \ac{AOA}, \ac{AOD}, and \ac{RSS}. 
Another localization method is \textit{proximity}, where a rough location of the \ac{UE} is determined by the location of some reference entities (e.g., \acp{BS} and anchors) in the proximity of the users. For instance, \ac{CID} and \ac{E-CID} techniques fall within this category.  

The positioning platforms can be classified into three main groups: \textit{(a) ad-hoc terrestrial}, \textit{(b) cellular}, and  \textit{(c) ad-hoc satellite} systems. 
In the following, we focus on the first two solutions. 
\begin{figure*}[t!]
 \centering
 \includegraphics[width=0.79\linewidth,draft=false]
 {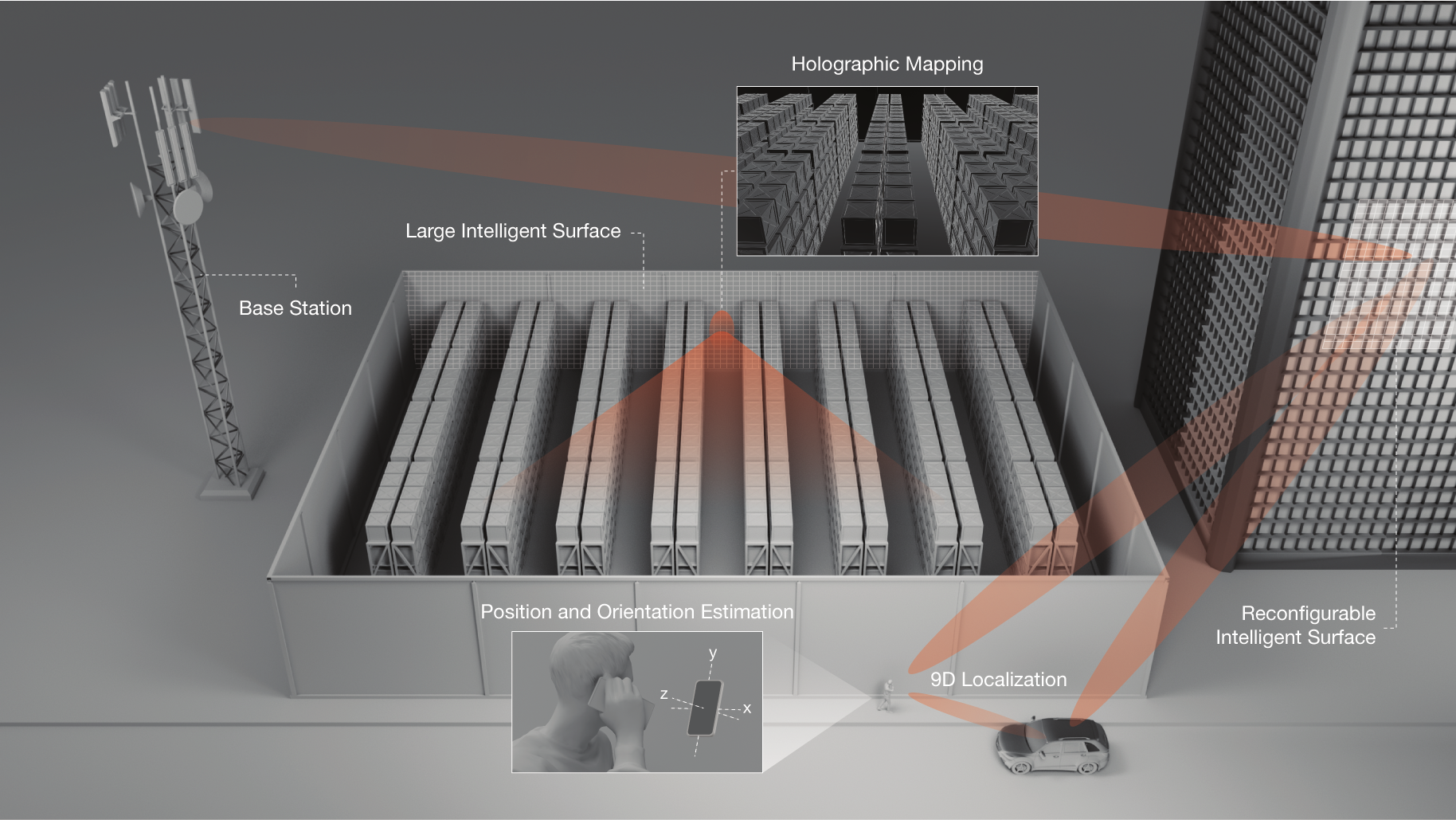}
 \caption{Pictorial vision of possible future communication, sensing and localization scenario.}
 \label{fig:fig1}
 \end{figure*}
\paragraph{Ad-hoc Positioning Systems} Different types of measurements can be processed to infer the user position, e.g.,  \ac{TOA}, \ac{TDOA}, \ac{AOA}, \ac{AOD}, and \ac{RSS} \cite{elzanaty2020reconfigurable}. 
Such measurements can be obtained through tailored technological solutions, spanning from ultrasounds, \ac{UWB}, \ac{mm-waves} towards visible light-based technologies, e.g.,  \ac{LiDAR} and \ac{VLP} technologies. To this purpose, in Table~\ref{tab:adhocloc} we summarize the main characteristics for each ad-hoc technology employed for wireless positioning, as well as the related achievable localization accuracy.

\paragraph{Cellular-based Positioning} Cellular networks are designed mainly to support high-quality communications. Only recently, communication and localization have been jointly designed and this integration will be further pushed in \ac{6G} mobile networks. In past generations, positioning was considered as an add-on service for communications whereas now cellular systems are also conceived as a mean to provide location information for users, covering a wider area compared to terrestrial ad-hoc techniques. To this purpose, Table~\ref{tab:cellularloc} displays the main features of cellular-positioning for various generations, till next \ac{6G}, including the related \ac{KPI}.
Indeed, next \ac{6G} is expected to dramatically impact the way wireless localization is conceived in cellular-based systems.
 \renewcommand{\arraystretch}{1.6}
\begin{table*}[t!]
 	\caption{Common ad-hoc localization techniques such as \ac{UHF}-\ac{RFID}, \ac{UWB}, \ac{mm-waves}, \ac{LiDAR}, and \ac{VLP}.}
 	 	\label{tab:adhocloc}
 	\centering
{\rowcolors{1}{gray!40}{orange!30}
\begin{tabular}{|p{3cm}|p{2.5cm}|p{2.5cm}|p{2.5cm}|p{2.5cm}|p{2.5cm}|}
	\hline 
	\rule[-1ex]{0pt}{2.5ex}  &  \ac{UHF}-\ac{RFID} & \ac{UWB}   & \ac{mm-waves}  & \ac{LiDAR} & \ac{VLP}  \\
	\hline 
	\rule[-1ex]{0pt}{2.5ex} Typical Frequency & 865-868 MHz &3.1-10.6 GHz  & 28, 60, 77 GHz  & 200 THz & 400-790 THz\\
	\hline
	\rule[-1ex]{0pt}{2.5ex} Typical Bandwidth & 200 kHz  & larger than 500 MHz  & larger than 2.1 GHz  & - &  150 MHz \\
	\hline
	\rule[-1ex]{0pt}{2.5ex}  Localization Technique   & \ac{RSS}, \ac{PDOA} & \ac{TOA}, \ac{TDOA} & \ac{TOA}, \ac{TDOA}, \ac{AOA}  & \ac{TOA} & \ac{RSS} \\
	\rule[-1ex]{0pt}{2.5ex} Coverage  & very small (less than 10~m) & moderate (less than 100~m) &   mode rate (less than 100~m)&  large (less than 500~m)  & tiny (less than 10~m)  \\
	\hline
	\rule[-1ex]{0pt}{2.5ex}  Accuracy &less than 5~m   & less than 10~cm  &  less than 1~cm  & less than 1~cm & less than 15~cm \\
	\hline
	\rule[-1ex]{0pt}{2.5ex} Cost  & very low  & moderate &   moderate/high &  high  &  very low\\
	\hline
	\end{tabular}
}
 \end{table*}
\renewcommand{\arraystretch}{1}
\renewcommand{\arraystretch}{1.6}
 \begin{table*}[t!]
	 	\caption{Cellular-based Localization Techniques, Features, Use Cases, and \ac{KPI} such as  \ac{OTDOA}, \ac{UAoA}, \ac{Multi-RTT} \ac{LCS}, \ac{PE}, \ac{LPP}, \ac{PRS}, \ac{NRPPa}, and \ac{ES}.}
	 		\label{tab:cellularloc}
	 	\centering
	 	{%
	{\rowcolors{1}{gray!40}{orange!30}
		\begin{tabular}{|p{3.5cm}|p{2cm}|p{2cm}|l|p{2.5cm}|p{3cm}|}
			\hline 
			\rule[-1ex]{0pt}{2.5ex}  &  \acs{2G} &\acs{3G}& \acs{4G} & \acs{5G}  & \acs{6G}  \\
			\hline 
			\rule[-1ex]{0pt}{2.5ex} Maximum Frequency & 1900 MHz &  2100 MHz &6 GHz  & 90 GHz  & 3 THz  \\
			\hline
			\rule[-1ex]{0pt}{2.5ex} Typical Bandwidth & 200 kHz  & 500 kHz  & 20 MHz  & 100 MHz  & 500 MHz   \\
			\hline
			\rule[-1ex]{0pt}{2.5ex}  Enhanced Techniques   & \ac{CID}, assisted \ac{TOA}, \ac{OTDOA}& \ac{CID}, assisted \ac{TOA}, \ac{OTDOA} &\ac{E-CID},  \ac{UTDOA} & {\ac{UAoA}, \ac{Multi-RTT},  \acs{NR} \ac{E-CID} }& Enhancing Previous Techniques (e.g., assisted with RIS and  AI) \\
			\hline
		    \rule[-1ex]{0pt}{2.5ex}  Localization Features &\ac{LCS}  & \ac{PE} &\ac{LPP}, \ac{PRS}  & \ac{NRPPa} & Joint Communication \& Sensing protocol  \\
		    \hline
				\rule[-1ex]{0pt}{2.5ex}  Use Cases  & \ac{ES}   &  \ac{ES}    &  \ac{ES}  & { \ac{ES},  industry, logistics, e-Health,  aerial}  & Holographic 9D Localization and Communications, Teleportation \\
			\hline
			\rule[-1ex]{0pt}{2.5ex} KPI horizontal positioning error outdoor (indoor) & less than 100 m, 67\%  & less than 100 m, 67\% & less than 50 m, 67\%  & less than 10 m (3 m), 80\%  &  less than 10 cm, 99.9\%  \\
			\hline
			\rule[-1ex]{0pt}{2.5ex} KPI vertical positioning error outdoor \& indoor  & NA   & NA & less than 3~m, 67\%  & less than 3~m, 80\%  & less than 10~cm, 99.9\%    \\
			\hline
		\rule[-1ex]{0pt}{2.5ex} Positioning latency & NA   & NA & NA  &less than 1 sec  & less than 0.1 sec    \\
			\hline
			\end{tabular}
		}
	}
\end{table*}
    \renewcommand{\arraystretch}{1}

\subsection{Applications}
{As previously highlighted, the advent of holographic radio will make available a larger number of degrees of freedom enhanced by the different operational \ac{EM} propagation regimes (e.g., in the Fresnel region). Moreover, the possibility to fully control the \ac{EM} field at both transmitter and receiver side, granted by the advent of new technologies, will help approaching the ultimate localization and communication performance limits.} 
Indeed, this will open the possibility to enable new applications, in light of next \ac{6G} scenarios. In fact, accurate positioning of mobile users can enhance the performance of communication systems, also called as location-aware communication. Some examples are:
\textit{(i)} geometric beams at the transmitters can be designed for  location-based beamforming without the need to estimate the full-band \ac{CSI}; \textit{(ii)} the system can adopt sophisticated spatial techniques to mitigate the interference; \textit{(iii)} location information and measured radio parameters over a long time period enable the construction of radio environment maps, opening many opportunities in terms of proactive resource allocation, without the need to know the instantaneous \ac{CSI}; \textit{(iv)} data can be communicated with low-latency by considering proactive location-based backhaul routing; \textit{(v)} the \ac{CSI} can be efficiently estimated from a minimal number of pilots by considering geometric channel models when the location of the user and the scatterers can be accurately computed; and \textit{(vi)} the radiation pattern of the antennas can be optimized to minimize the users' exposure to \ac{EMF} based on the orientation and location estimation of the \ac{UE}.

We now present some potential applications that require high-accuracy localization and low-latency. In Fig.~\ref{fig:fig1}, an example of holographic localization scenario enabled by \acp{LIS} is depicted. 
In this scenario, the intelligent surface in outdoor acts as a passive reflector for assisting the \ac{BS} in estimating the  9D location (position, orientation, and speed) of the user and vehicle. The indoor intelligent surface in the warehouse can be active and it allows holographic localization and mapping.
Towards this vision, some use cases are reported in the following.

\paragraph{Industrial Internet of Things}
Accurate positioning is crucial for exploiting the full potentials of sensors and actuators in industrial \ac{IoT} scenarios. All {\em things} (e.g., sensors, machines, tools, or humans) can be equipped with tags wirelessly connected to provide communication services and high accuracy positioning.  
For example, mobile robots require accurate information about their location and the surrounding environment to properly navigate through congested warehouses.  
\paragraph{Simultaneous
Wireless Information and Power Transfer} \ac{6G} are expected to foster systems equipped with near-pencil beam antenna arrays and high directivity. Hence, there will be the possibility not only to proficiently transfer information through the generated beams, but also to  energize the devices through the same communication link. Localizing with extremely high accuracy is demanded to properly focus the power transfer and reduce the amount of wasted energy.  

\paragraph{Intelligent Transportation Systems}
Accurate positioning can help in enabling better traffic routing as well as in the navigation support of autonomous cars and drones through the avoidance of collisions enabling services such as  autonomous flying taxis.  
\paragraph{\Ac{XR}}
Most of \ac{XR}-based systems require accurate positioning. For example, one application permits the users to modify the urban design by creating and removing parts of existing buildings. 
The reproduced illusion requires high localization and orientation-detection accuracy to provide the user with an immersive experience. 
%

%


%
\section{Holographic Localization Enabling Technologies}
\label{sec:ris}

To enable holographic localization, we need technologies to realize electrically large antennas and reflectors so that localization occurs in the near-field propagation regime, thus enabling holographic localization. Moreover,  such technologies should allow to achieve a full control of the \ac{EMF}, thus moving some of the signal processing from the digital to the \ac{EM} level, achieving gains in terms of flexibility,  lower latency, power consumption, and complexity.


Conventional discrete antenna solutions entail the adoption of antenna elements usually spaced apart by half-wavelength. In this case, each cell is equipped with an antenna and acts as an independent unit that modifies the behaviour of the wave (that can be also reflected) in a desired manner (beam-tailoring capability). Nevertheless, such a technique requires a large number of  \ac{RF} chains for the increased number of antenna elements, leading to higher cost and bulky implementations. Also, with a half-wavelength separation between antennas,  if we increase the operating frequency while fixing  the number of antenna elements, the users will fall easily in the far-field of the array with planar waves. Since we lose information about the curvature of the wavefront,  it becomes harder to jointly estimate the range and \ac{AOA}, leading to poor localization performance.

In the following, we discuss potential techniques that can enable holographic localization by allowing more users to be in the near-field of the array and exploiting the entire position information associated with the propagating \ac{EM}. 

\subsection{Extremely Large Aperture Arrays}
Massive \ac{MIMO} with considerably large distance between antenna elements (compared to the wavelength) can enable the realization of  \ac{ELAA}. These arrays have significantly high spatial resolution, enabling accurate positioning performance due to their large apertures compared to arrays with the same number of antennas, but with smaller apertures. 

Moreover, buildings coated with \ac{ELAA} will extend the near-field region even for several kilometers around the array, allowing to manage more degrees of freedom which can be exploited to enhance the localization. For instance, since the
\ac{UE} can be likely located in the near-field of an \ac{ELAA}, it will be possible to directly retrieve the distance (ranging) information from the spherical wavefront together with the
\ac{AOA} information \cite{AmiriHeath:18}.
%
In addition, antennas will experience a different \ac{RSSI} with great probability due to the large distance between them (e.g., imagine two elements at different corners of a building), so that further location information can be gathered. In addition,
it might happen that, due to the large aperture, some antennas of the \ac{ELAA} might be in \ac{NLOS} with the \ac{UE}, whereas others
might keep the visibility with it.
%
Therefore, the study of the visibility region of the \ac{ELAA} is of paramount importance for localization.

Another benefit is that in near-field the signal can be focused on a point, i.e., the user location, rather than a direction. If we have a rough initial estimation for the user's location,  the received \ac{SNR} at the user can be boosted, the interference to other users is reduced, and a more accurate user's position can be estimated. 


Nevertheless, several challenges still need to be properly addressed, such as the presence of strong phase ambiguities that could dramatically affect the localization accuracy, as well as the requirement of a tight synchronization among the antennas within the \ac{ELAA}.




\subsection{Intelligent Surfaces}
 The introduction of metamaterials, used to realize intelligent surfaces for coating both small and large objects, seems a viable and promising solution for the realization of holographic localization.\footnote{Metamaterials are materials capable of offering properties not available in nature.} In this way, these surfaces become controllable (i.e., reconfigurable) and intelligent (i.e., able to focus the power towards the desired targets).


 
Intelligent surfaces can be employed in different functional modes: \textit{(i)} in transmission by modulating the phases of the radiating elements and/or in reception when they are equipped with a set of \ac{RF} chains; \textit{(iii)} in reflection when they are used as a (passive) relay of multipath components and it controls the reflections in real-time. 
In the following, we will discuss  these modes in  details.
\subsubsection{Intelligent Surfaces as Active Antennas (LIS)}
\label{sec:txrxmode}
\Acp{LIS} can be used as transmitter, receiver or for both purposes. 
In this scenario, they can be interpreted as compact and low cost antenna arrays that might be exploitable at the \ac{BS}. Such antennas can be fabricated with either tiny antennas or metamaterial radiating elements capable to actively communicate over a wireless link \cite{shlezinger2021dynamic}. When \acp{LIS} are used actively and with the adoption of many tiny antennas, they represent a natural evolution of massive \ac{MIMO} technology \cite{SangMowWin:21}.
%

{Recently, it has been investigated the possibility to adopt dynamic metasurfaces as active antennas
to control the transmit/receive beam patterns with advanced hybrid A/D signal processing capabilities  \cite{shlezinger2021dynamic}.} 
Compared with conventional antennas implemented by patch antennas and phase shifters, dynamic metasurface antennas operate at low power consumption and cost, while naturally implementing \ac{RF} chain reduction without requiring dedicated analog circuitry but by exploiting metamaterials located within wave-guides.

{Another possibility is to place a reconfigurable \ac{EM} lens in front of a single transmitter/receiver antenna, such that part of the signal processing is performed in analog through large surfaces.} 
In this way, there is the advantage that the digital signal processing is dramatically reduced, as it is delegated to the lens that operates in the analog domain. On the other side, the cost to be paid is a reduced flexibility, as only one \ac{RF} chain is connected to the antenna that collects the resulting signal after the lens processing. 

\subsubsection{Intelligent Surfaces as Reflectors (RIS)}
\label{sec:reflectmode}
\acp{LIS} can be also employed to ease and enable the wireless localization between the \ac{BS} and multiple users by acting as a mean to control the multipath, and they are also referred to as \acp{RIS} or \acp{DCS}, since each element of the surface is treated as local scatterer. 
Indeed, this is in principle different from relays that are usually active or they require more energy and sophisticated processing operations.
{Indeed, this emerging technology exploits the possibility to dynamically and artificially adjust the physical properties, such as permittivity and permeability, of the \ac{EM} waveforms to obtain some desired electrical or magnetic characteristics that in principle are not available in nature \cite{Huang2020holographic}. Operating like this, radio waves can be shaped in a way that the reflected signals might not obey the Snell's law, but rather a generalized Snell's law.} 
{As an example, such an effect can be realized with} unit cells made of metallic or dielectric patches that can be modeled as a passive scattering element. To preserve the energy and cost efficiency, each unit cell can equipped with some low power tunable electronic circuits, e.g., PIN diodes or varactors, and sometimes there is the add-on of sensing elements. 
{An alternative is represented by metaprism, which refers to a passive and non-reconfigurable metasurface that acts as a metamirror.} Its reflecting properties are frequency-dependent within the signal bandwidth and they can be optimized at the \ac{BS} by proper frequency resource allocation to increase the communication and localization coverage even in situations where the \ac{LOS} is obstructed. 

\section{Performance Limits and Enabling Algorithms}\label{sec:loc}

As previously mentioned, {\em holographic localization} is the capability to fully exploit the \ac{EM} characteristics of the incident wavefront to infer the position information. In other words, it refers to the possibility of recording a quasi-continuum measurement profile through which the position and the orientation of a user are inferred.
For example, when using an array of antennas or a metasurface whose dimension is large enough to consider the surrounding users in the near-field region, the phase profile of the impinging waveform provides sufficient information to estimate the positions. In fact, the plane wave approximation is no more valid in the near-field and the spherical characteristic of the \ac{EM} wave brings all the needed information for the positioning, i.e., distance and angle information \cite{guerra2021near}. 
To this aim, in the following, we illustrate  \textit{(i)} the latest contributions on establishing the localization performance limits in \ac{RIS}-aided scenarios; \textit{(ii)} some of the localization algorithms and their complexity. 
\subsection{Fundamental Limits} The localization performance limits provide a lower bound on the achievable estimation mean square error of any (unbiased) localization estimator. Such limits, typically based on \ac{CRLB}, represent benchmarks for practical estimators and depend on various parameters, such as, for example, the adopted technology, the geometric scenario, the array geometry, the presence/absence of a LOS link, the accuracy and number of measurements, and the presence/absence of any prior information on parameters. Moreover, any type of collaboration between nodes should be taken into account by these limits as well as the adopted waveforms and codebooks, and the presence of synchronization and technological impairments. A survey on the \ac{CRLB} for classical positioning problems can be found in \cite{win2018theoretical} 
and, only recently, research on fundamental limits has advocated the importance of considering near-field propagation induced by the use of large surfaces where the ultimate localization and orientation fundamental limits are derived in scenarios with a \ac{RIS} \cite{hu2018beyond,elzanaty2020reconfigurable} or with a \ac{LIS} \cite{guerra2021near,SangMowWin:21}.  
More specifically, in the derivation of the near-field \ac{CRLB} for localization, authors in \cite{SangMowWin:21} put together estimation theory and wave propagation, accounting for the
electromagnetic field over a defined region.
Indeed, the aforementioned near-field contributions have shown a performance enhancement in terms of positioning accuracy and coverage when using \acp{RIS} and the possibility of using a single node and narrow-band signals for localization. Therefore, they can be considered as a first step towards the concept of holographic positioning. 
\begin{figure}[t!]
	\centering
	\includegraphics[width=0.75\linewidth,clip]{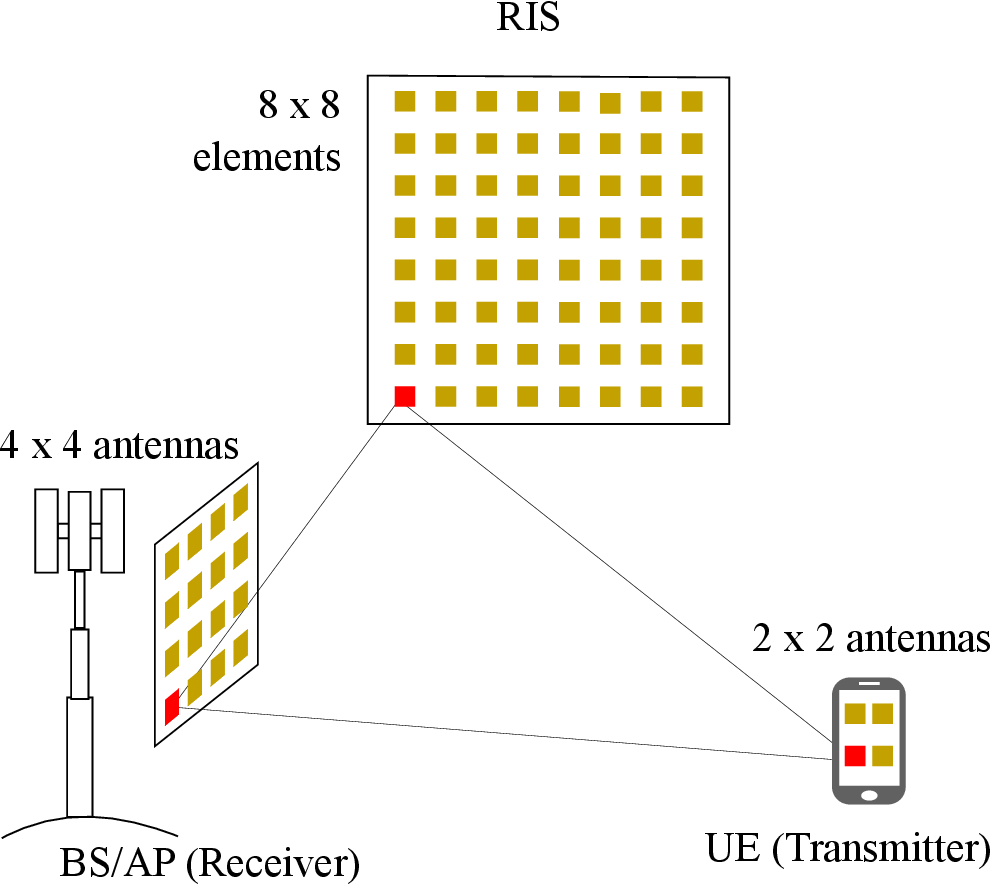}
	\caption{Considered localization scenario.}
	\label{fig:scenario}
\end{figure}

\subsection{RIS-aided Localization Approaches} 
Localization algorithms assisted by \acp{RIS} usually are based on the assumption that a discrimination between the direct path, representing the link between the user-\ac{BS}, and the reflected path controlled by the \ac{RIS} is possible, for example by adopting CSI estimation approaches 
\cite{JinWymJun:J21}. Usually, this entails accumulating a large number of pilot symbols and adopting random phase configurations.  Subsequently, both direct positioning or two-step localization approaches can be used to localize the user. 
For example, in \cite{zhang2020positioning} where a dual \acp{RIS} scenario is considered, the delay difference of the direct and reflected paths corresponding to each \ac{RIS} present in the environment is calculated via the cross-correlation of the signals over the two paths and the location of the user is finally estimated according to the geometric laws.
In \cite{ma2020indoor} an indoor positioning application is considered and, thus, the \ac{UWB} technology is proposed in conjunction with the adoption of \ac{RIS} for its ability to resolve multipath. On the other hand, the near-field is exploited in \cite{RinchiElzanaty:22} to estimate the ranges, \acp{AOA}, and \acp{AOD} for the \ac{UE}-\ac{RIS} and \ac{RIS}-\ac{BS} links using a sparse recovery algorithm.
%

The performance of the aforementioned localization algorithms depends heavily on the \ac{RIS} phase profile. In fact, the phase shift induced at each \ac{RIS} element should be adequately designed to improve the localization performance, which is obtained as a solution to an optimization problem. Several performance metrics can be considered as objective functions that are optimized to improve the positioning performance of such as the received \ac{SNR}, \ac{CRLB} on location and orientation estimation, and algorithm-tailored localization errors \cite{elzanaty2020reconfigurable}.

\section{Case Study}


In this case study, we focus on uplink positioning which is an important add-on service in 6G communication systems. In particular, we consider a single-anchor localization scheme in \ac{RIS}-assisted environment, where a \ac{BS} estimates the location and orientation of a \ac{UE} from the direct and \ac{RIS}-reflected signals. The \ac{BS} can exploit the full {\em holographic} profile of the impinging signal, as we account for the wavefront curvature by considering the near-field model rather than the far-field approximation that assumes planar wavefront. The system operates with a center frequency of 28 GHz. 
%
The \ac{UE} transmits an \ac{OFDM} signal with -10 dBm allocated power per subcarrier. The \ac{CRLB} of this systems is derived in the recent paper \cite{elzanaty2020reconfigurable}.\footnote{The numerical results are conducted using \textsc{MATHEMATICA}\textsuperscript{\textcopyright}, and the source code is available at \url{https://tinyurl.com/yc6mev4d}.}

{To corroborate our vision, we report in Fig.~\ref{fig:scenario} an example of the scenario of interest, where the \ac{RIS} location is in (5,5,1) meters, while the \ac{BS} is in the center.
%
%
%
%
%
\begin{figure}[t!]
	\centering
	\includegraphics[width=1\linewidth,clip]{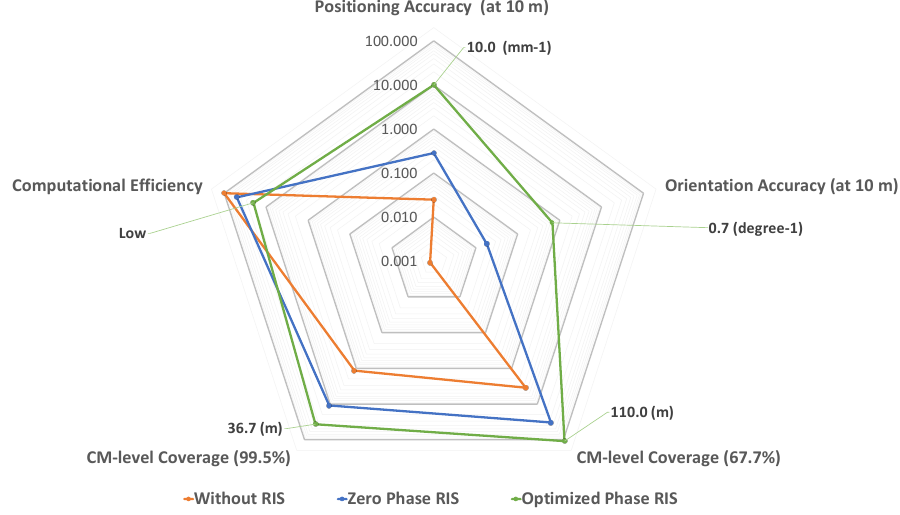}
	\caption{Radar chart on the localization performance  for  various architectures without and with  \acp{RIS}, according to the scenario displayed in Fig.~\ref{fig:scenario}.}
	\label{Fig:PhaseDesignFig}
\end{figure}
Then, in Fig.~\ref{Fig:PhaseDesignFig}, we consider the performance for such a scenario accounting for the following schemes:}
\textit{i) Without \ac{RIS}:} single-anchor localization without the help of a \ac{RIS};
\textit{ii) Zero Phase RIS:}  the \ac{RIS} acts as an \ac{EM} mirror obeying the Snell's law, i.e., the phase shifts are all zero; 
\textit{iii) Optimized Phase RIS:} the \ac{RIS} is designed to minimize the  \ac{CRLB} on position or orientation estimation, i.e.,  \ac{PEB} or \ac{OEB}.
The performance is indicated in terms of several metrics, i.e., 
\textit{i) positioning/orientation accuracy:} the reciprocal of the \ac{PEB}/\ac{OEB} obtained from the \ac{CRLB} computed at a distance of 10~m from the \ac{BS};
\textit{ii) CM-level Coverage (67.7\% or 99.5\%):} the maximum distance from the \ac{BS} such that we are sure by 67.7\% or 99.5\% that the localization error is less than 1~cm;
\textit{iii) Computational efficiency:} the reciprocal of the computational complexity.  We set the number of \ac{UE}, \ac{BS}, and \ac{RIS} antennas to 4, 16, and 64, respectively.  We can see that the optimized \ac{RIS} phase leads to higher position accuracy, orientation accuracy, and coverage compared to the without-\ac{RIS} scheme by factors of about 400~\!\!\textrm{x}, 812\!\!~\textrm{x}, and 31~\!\!\textrm{x}, respectively, at the expense of an increased computational complexity.


\begin{figure}[t!]
	\centering
	\vspace{-1.37cm}
	\includegraphics[width=1\linewidth,clip]{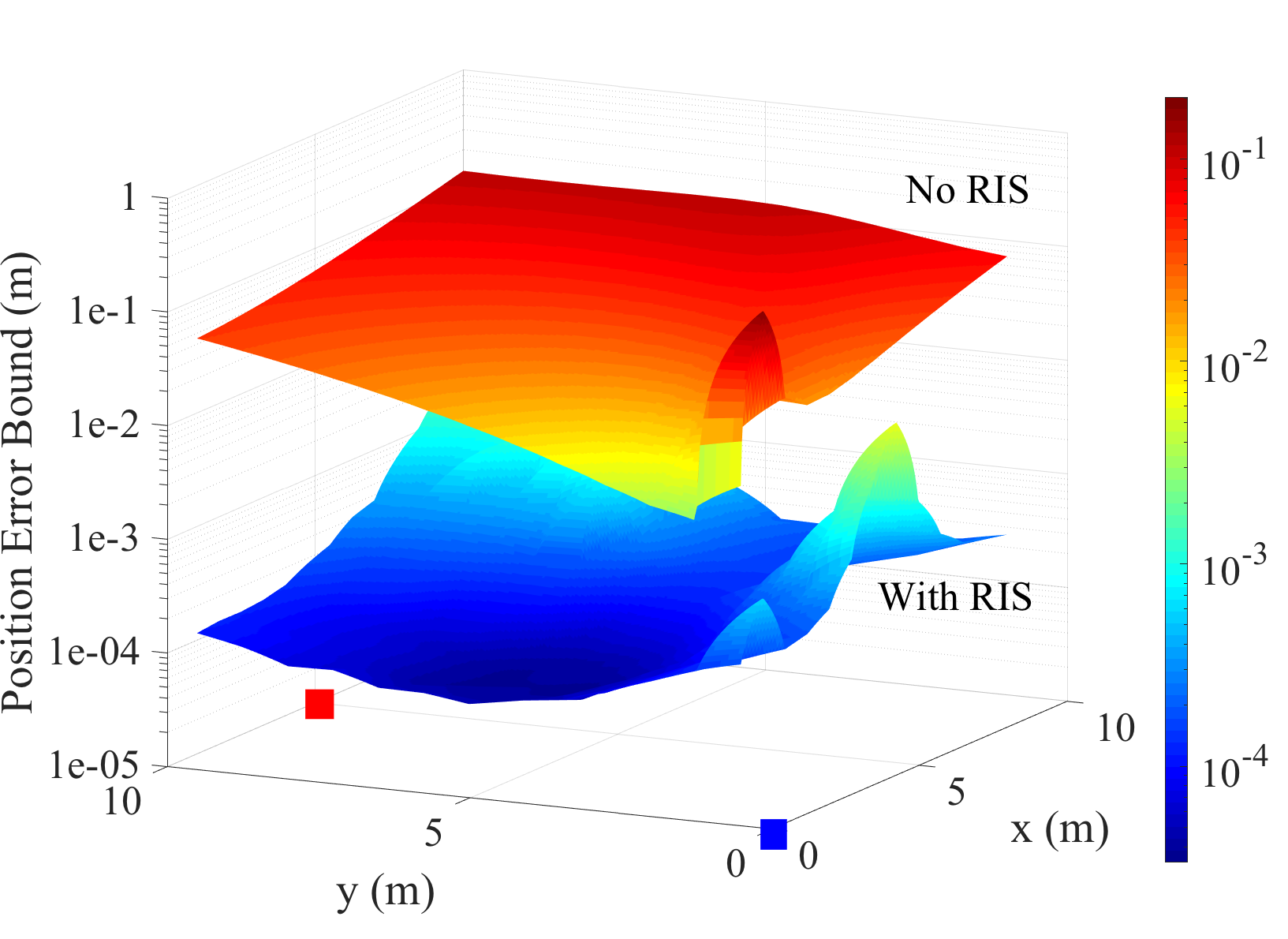}
	\caption{The \ac{PEB} vs various \ac{UE} location for scenarios with and without \ac{RIS}, with blue and red markers indicating the \ac{BS} and the \ac{RIS} position, respectively.}
	\label{fig:PEB3D}
\end{figure}

The impact of RIS-assisted environment for various \ac{UE} locations is also evidenced in the 3D maps of Fig.~\ref{fig:PEB3D}. We can see that the error is minimal when the \ac{UE} is near the \ac{RIS}, which is located at (5,10,1) meters, with a localization error that is almost three orders of magnitude less than the scheme without \ac{RIS}. The \ac{RIS} plays in a twofold manner: it creates a new link between the \ac{BS} and the \ac{UE}, and it expands the near-field region, as the \ac{UE} can be in the near-field of the \ac{RIS} but in the far-field with the \ac{BS}.
These numerical considerations represent the first step towards
holographic localization, showing that it is possible to attain a
localization accuracy in the order of the millimeter, and an orientation accuracy of a fraction of degree. Moreover, this put in evidence the need of having a properly \ac{RIS} phase design, which is still a
challenging when \ac{RIS}-assisted localization is performed. 

\vspace{-0.3cm}
{\section{Conclusion and Future Directions}}

In this paper, we have motivated our vision towards holographic localization, enabled by \acp{LIS}, as an essential feature for next wireless networks. {There are several key research aspects that present unsolved issues,} but the road ahead is promising, and it is expected to nurture new challenging studies towards the establishment of \ac{6G}. 
In the following, some of the future directions and challenges for holographic localization with intelligent surfaces are stated.



\textbf{High Complexity Iterative Algorithms}
The problem formulation of the \ac{RIS} phase design usually requires an initial rough position estimate for the device that needs to be localized. Therefore, the localization problem often involves multiple iterations, where the estimated location of the device in the last iteration is used to compute the \ac{RIS} phases for the next iteration. 
This process should be repeated iteratively until the algorithm converges to the actual location. Therefore, there is a great challenge of obtaining the position information with low overhead and extremely low latency.

\textbf{Multi-Objective Function in 9D Localization}
Some \ac{6G} use cases, such as augmented reality based applications, require accurate 9D localization, i.e.,  the position, orientation, and speed estimation errors should be minimized. 
Unfortunately, in general it is challenging to optimize the \ac{RIS} phases such that all the aforementioned errors are simultaneously minimized, as required for 9D localization. Hence, the phase design turns into a multi-objective optimization problem, requiring more sophisticated algorithms.
In this sense, efficient solutions for the \ac{RIS} phase design by optimizing multi-objective functions are still lacking. 

\textbf{Ad-hoc Waveform Design}
In general, for 9D localization, multiple waveforms can also be designed for different time slots in order to achieve better accuracy for the estimation of different parameters in a time division fashion.

Moreover, following a trend started by LTE Advanced, the 6G waveforms will be designed for joint communication and  localization purposes. In this regard, the joint design of the \ac{RIS} phases and the  beamforming codebooks at the \ac{BS} and \ac{UE} is essential. Moreover, a distinction should be made according to the use of \ac{RIS} either being a transmitter or a reflector. The joint optimization problem can be also challenging, increasing the complexity of the localization algorithms.  

\textbf{Holographic Simultaneous Localization and Mapping} \ac{6G} can exploit holographic radio to increase the ambient awareness by reconstructing the surrounding environment (\emph{holographic mapping}) and by allowing a user to self-localize with respect to the reconstructed map. In this perspective, {holographic} \ac{SLAM} will allow user to improve their perception of the environment and their interaction with it. 

\textbf{AI for Holographic Localization}
Machine learning approaches can be used to assist in solving the optimization problem of the phase profile.
In addition, the user's location and orientation can be inferred from the received signal with machine learning approaches, where a deep neural network can be trained by mapping the environment through a sub-sample of random location for the UE. Nevertheless, the required training data for data-driven schemes with machine learning can be massive to achieve the target accuracy. In this regard, another machine learning method that does not necessitate a large amount of possible \ac{UE} locations is more relevant.



\bibliographystyle{IEEEtran}
\bibliography{IEEEabrv,Elzanaty_bibliography.bib,LISBIB.bib}


\medskip

{\bf Ahmed Elzanaty} (S'13-M'18-SM'22) received the Ph.D. degree  in Electronics, Telecommunications, and Information technology from the University of Bologna, Italy, in 2018.  
He is a lecturer at the Institute for Communication Systems (ICS), University of Surrey, UK. 
His research interests include wireless localization, coded modulation, and compressive sensing. 

\medskip

{\bf Anna Guerra} (M'16) received the  Ph.D. degree in Electronics, Telecommunications, and Information Technology from the University of Bologna (UNIBO), Italy, in 2016. She is a Research Fellow at UNIBO.
From 2018 to 2020, she was a  Marie Curie Fellow  with UNIBO and Stony Brook University, New York, USA. Her research interests include wireless sensor networks, radio localization, and  signal processing. She is an Associate Editor for IEEE Communications Letters. 

\medskip

{\bf Francesco Guidi} (M'13) received the Ph.D. degree in electrical engineering from the University of Bologna and the Ecole Polytechnique, Paris.
He is now a Researcher at the IEIIT-CNR, Italy. From 2015 to 2017 he was a Marie Curie Fellow at CEA-LETI, France. His research interests include radar networks, radio localization, and multi-antenna systems.

\medskip

{\bf Davide Dardari} (M'95–SM'07) received the M.S. and Ph.D. degrees in electrical engineering from the University of Bologna. He is a Full professor at the
University of Bologna and CNIT, Italy.
Since 2005, he has been a research affiliate at the Massachusetts Institute of Technology, Cambridge. His research interests are
wireless communications, localization techniques, and
distributed signal processing. 

\bigskip

 {\bf Mohamed-Slim~Alouini} (S'94-M'98-SM'03-F'09)  was born in Tunisia. He received the Ph.D. degree in Electrical Engineering from the California Institute of Technology (Caltech), USA, in 1998. He served as a faculty member in the University of Minnesota, USA, then in the Texas A\&M University, Qatar, before joining King Abdullah University of Science and Technology (KAUST), Saudi Arabia, as a Professor of Electrical Engineering in 2009. His current research interests include modeling, design, and performance analysis of wireless communication systems.





\end{document}